\documentstyle[12pt]{article}

\newcounter{mine}
\begin{document}

\begin{center}
{\large {\bf QUADRATIC SOLITONS IN CUBIC CRYSTALS}} \vspace{1cm}

\bigskip

Boris V. Gisin\footnote{%
e-mail address: gisin@eng.tau.ac.il}

\smallskip

{\it Department of Electrical Engineering - Physical Electronics, Faculty of
Engineering, Tel-Aviv University, Tel-Aviv 69978, Israel}

\bigskip

Boris A. Malomed\footnote{%
e-mail address: malomed@eng.tau.ac.il}

\smallskip

{\it Department of Interdisciplinary Studies, Faculty of Engineering, Tel
Aviv University, Tel Aviv 69978, Israel}

\bigskip

\abstract
\end{center}

Starting from the Maxwell's equations and without resort to the paraxial
approximation, we derive equations describing stationary (1+1)-dimensional
beams propagating at an arbitrary direction in an optical crystal with cubic
symmetry and purely quadratic ($\chi ^{(2)}$) nonlinearity. The equations
are derived separately for beams with the TE and TM polarizations. In both
cases, they contain $\chi ^{(2)}$ and cubic ($\chi ^{(3)}$) nonlinear terms,
the latter ones generated via the cascading mechanism. The final TE
equations and soliton solutions to them are quite similar to those in
previously known models with mixed $\chi ^{(2)}$-$\chi ^{(3)}$
nonlinearities. On the contrary to this, the TM model is very different from
previously known ones. It consists of four first-order equations for
transverse and longitudinal components of the electric field at the
fundamental and second harmonics. Fundamental-soliton solutions of the TM
model are also drastically different from the usual $\chi ^{(2)}$ solitons,
in terms of the parity of their components. In particular, the transverse
and longitudinal components of the electric field at the fundamental
harmonic in the fundamental TM solitons are described, respectively, by odd
and single-humped even functions of the transverse coordinate. Amplitudes of
the longitudinal and transverse fields become comparable for very narrow
solitons, whose width is commensurate to the carrier wavelength.

\newpage

\section{Introduction}

Optical crystals with the cubic symmetry are the basis of contemporary
photonics. In particular, spatial optical solitons generated by the $\chi
^{(2)}$ (quadratic) nonlinearity, see reviews \cite{YSK,Jena}, are possible
in these media. In the simplest case, these are two-component pulses
composed of fundamental-frequency (FF) and second-harmonic (SH) waves.

Usually, solitons in noncentrosymmetric crystals are considered when
directions of the light propagation and polarization coincide with the
crystal axes. However, in cubic crystals it is possible to investigate
solitons carried by a beam which is propagating in {\em arbitrary}
direction, the situation being similar to that known in electrooptics \cite
{Gisin1}. In the present work, our objective is to study spatial solitons of
both TE and TM (transverse-electric and transverse-magnetic) types with the
arbitrary propagation direction of the carrier wave. First, we derive a
corresponding general model (actually, we obtain a system of ODEs to
describe a stationary shape of the (1+1)-dimensional solitons). The model is
derived directly from the Maxwell's equations, without using the paraxial
approximation, so that the final equations apply as well to very narrow
(subwavelength) spatial solitons.

In the case of the TE polarization, which is considered in section 2, the
obtained model differs from the well-known simplest model giving rise to the 
$\chi ^{(2)}$ solitons by a single additional {\em cubic} ($\chi ^{(3)}$)
term in the equation for the transverse FF component, which is generated by
the underlying quadratic nonlinearities via the cascading mechanism (see a
review \cite{Torner}). As solitons in this TE model are very close to those
in the known $\chi ^{(2)}$-$\chi ^{(3)}$ models, we consider them, also in
section 2, only briefly.

Unlike the TE case, the final model for the TM solitons, which is derived in
section 3, turns out to be essentially novel, consisting of four first-order
equations. Two equations for transverse FF and SH electric fields contain
only quadratic nonlinear terms, while two other equations, for longitudinal
electric fields, contain cascading-generated cubic terms too. Then, in
section 4 we find, in a numerical form, stationary solutions for both
fundamental and higher-order solitons in the TM model. The fundamental TM
solitons turn out to be drastically different, as concerns the parity of
their components, from solitons in the traditional $\chi ^{(2)}$ models:
their longitudinal FF electric field, which has no counterpart among
components of the usual $\chi ^{(2)}$ solitons, is described by an even
single-humped function, while the transverse TE electric field, which is an
analog of the FF field in the usual solitons, is an {\em odd} function of
the transverse coordinate $x$. Moreover, the transverse SH electric field,
although being an even function, is not monotonically decaying in the
interval $0<x<\infty $, but rather changes its sign at a finite $x$.

The $\chi ^{(2)}$ contribution to the polarization induced by the electric
field with {\em real} vectorial components $E_{n}$ in noncentrosymmetric
dielectric crystals can be represented in the form $P^{(2)}=d_{lmn}E_{m}E_{n}
$ (summation over repeated subscripts is assumed), where $d_{lmn}$ is a
nonlinear susceptibility tensor \cite{Tensor}. Obviously, it is symmetric
with respect to the permutation of the subscripts $m$ and $n$. 
For cubic noncentrosymmetric crystals in the crystollagraphic coordinates $%
\widetilde{x}$,$\widetilde{y}$,$\widetilde{z}$, nonvanishing components\ of
the tensor are $\widetilde{d}_{123}=\widetilde{d}_{213}=\widetilde{d}%
_{312}\equiv d$ \cite{Tensor}.

We will consider light propagation in an arbitrary direction, relative to
the axes $\widetilde{x}$,$\widetilde{y}$,$\widetilde{z}$, in a new
coordinate frame $(x,y,z)$, the $z$-axis being aligned with the propagation
direction. The new frame may be generated by consecutive rotations of the
initial one by angles $\varphi ,\;\vartheta $ and $\gamma $ around the $%
\widetilde{z}$, $x$, and $z$ axes, respectively. The matrix $\left(
a_{mn}\right) $ of the orthogonal transformation to the new frame, $%
x_{k}=a_{kl}\widetilde{x}_{l}$, where $\widetilde{x}_{k}$ and $x_{k}$ stand
for the sets $(\widetilde{x},\widetilde{y},\widetilde{z})$ and $(x,y,z)$, is 
\begin{equation}
\left( 
\begin{array}{ccc}
\cos \varphi \cos \gamma -\cos \vartheta \sin \varphi \sin \gamma  & \sin
\varphi \cos \gamma +\cos \vartheta \cos \varphi \sin \gamma  & -\sin
\vartheta \sin \gamma  \\ 
-\cos \varphi \sin \gamma -\cos \vartheta \sin \varphi \cos \gamma  & -\sin
\varphi \sin \gamma +\cos \vartheta \cos \varphi \cos \gamma  & -\sin
\vartheta \cos \gamma  \\ 
-\sin \vartheta \sin \varphi  & \sin \vartheta \cos \varphi  & \cos
\vartheta 
\end{array}
\right) \,.  \label{a}
\end{equation}
Components of the dielectric displacement vector in the new frame are 
\begin{equation}
D_{k}=\varepsilon E_{k}+2d_{kmn}E_{m}E_{n},  \label{D'}
\end{equation}
where $\varepsilon $ is the dielectric constant, and 
\[
d_{kmn}=a_{kl}a_{mp}a_{nq}\widetilde{d}_{lpq}\equiv
2d(a_{k1}a_{m2}a_{n3}+a_{k2}a_{m1}a_{n3}+a_{k3}a_{m1}a_{n2})
\]
is the $\chi ^{(2)}$ susceptibility tensor in the new frame. In the general
case, all the 27 tensor components are different from zero, but they depend
only on one constant $d$ and three angles $\varphi ,\;\vartheta $ and $%
\gamma $. 

Below, the full electric and magnetic fields, including both FF and SH
components, will be sought for in the form 
\begin{eqnarray}
E_{k} &=&a_{k}\cos \left( 2\Phi \right) +b_{k}\sin \left( 2\Phi \right)
+A_{k}\cos \Phi +B_{k}\sin \Phi ,  \label{E} \\
H_{k} &=&p_{k}\cos \left( 2\Phi \right) +q_{k}\sin \left( 2\Phi \right)
+P_{k}\cos \Phi +Q_{k}\sin \Phi .  \label{H}
\end{eqnarray}
Here, the FF phase is $\Phi =\beta z-\omega t$,$\;\omega $ is the frequency, 
$t$ is time, and $\beta $ is the propagation constant, the small and capital
letters standing for the SH and FF amplitudes, respectively. In the general
case, with the carrier wave propagating in an arbitrary direction, solutions
are of the mixed TE-TM type, i.e., they have all the electric and magnetic
field components. However, pure TE and TM modes will be found below for
certain angles $\varphi ,\;\vartheta $ and $\gamma $, including nontrivial
cases when the propagation and polarization directions are {\em not} aligned
with the crystallographic axes ($\widetilde{x},\widetilde{y},\widetilde{z}$).

To conclude the introduction, it is relevant to mention several other models
of the $\chi ^{(2)}$ type which take into regard the polarization of the
fields. They describe bimodal fields in an isotropic medium \cite{Boardman},
or the so-called type-II interactions between the FF and SH waves, with a
single (see, e.g., Refs. \cite{single}) or double \cite{double} parametric
resonance between them. However, none of these previously studied models
included longitudinal components of the fields, nor magnetic field was
explicitly introduced in them.

\section{The TE case}

First, we consider a simpler case with the FF wave of the TE type, i.e.,
when the electric FF field has only one component $A_{2}$, which is a
function of $x$, but directed along the $y$ axis. Then, the only nonzero FF
components are $A_{2}\equiv A$,$\;P_{1}\equiv P$,$\;$and $Q_{3}\equiv Q$,
the latter one being longitudinal magnetic field. Substituting $E_{k}$ and $%
H_{k}$ from Eqs. (\ref{E}) and (\ref{H}) into the Maxwell's equations, it is
possible to eliminate the magnetic-field components and obtain a set of
differential equations for the electric components only: 
\begin{eqnarray}
A^{\prime \prime } &=&\left( \beta ^{2}-\beta _{1}^{2}\right) A-2\beta
_{0}^{2}A(d_{221}a_{1}+d_{222}a_{2}+d_{223}a_{3}),  \nonumber \\
a_{1}^{\prime \prime } &=&4\left( \beta ^{2}-\beta _{2}^{2}\right)
a_{1}-4\beta _{0}^{2}d_{122}A^{2}-\varepsilon
_{2}^{-1}d_{122}(A^{2})^{\prime \prime },  \nonumber \\
a_{2}^{\prime \prime } &=&4\left( \beta ^{2}-\beta _{2}^{2}\right)
a_{2}-4\beta _{0}^{2}d_{222}A^{2},  \nonumber \\
a_{3}^{\prime \prime } &=&4\left( \beta ^{2}-\beta _{2}^{2}\right) \left(
a_{3}+\varepsilon _{2}^{-1}d_{322}A^{2}\right) ,  \nonumber \\
\,\,\,b_{1}^{\prime \prime } &=&4\left( \beta ^{2}-\beta _{2}^{2}\right)
b_{1}-2\beta \varepsilon _{2}^{-1}d_{322}(A^{2})^{\prime },  \nonumber \\
b_{3}^{\prime \prime } &=&4\left( \beta ^{2}-\beta _{2}^{2}\right)
b_{3}+2\beta \varepsilon _{2}^{-1}d_{122}(A^{2})^{\prime },  \label{eqTE} \\
b_{2}^{\prime \prime } &=&4\left( \beta ^{2}-\beta _{2}^{2}\right) b_{2},
\end{eqnarray}
where the prime stands for $d/dx$, $\beta _{0}\equiv \omega /c$ is the FF
propagation constant in vacuum, $c$ being the light velocity in vacuum, $%
\;\beta _{1,2}^{2}\equiv \varepsilon _{1,2}\beta _{0}^{2}$, and $\varepsilon
_{1,2}$ are FF and SH dielectric permeabilities. From the condition that the
fields must vanish at $x=$ $\pm \infty $, it immediately follows from the
last equation in the system (\ref{eqTE}) that $b_{2}\equiv 0$, and it can be
shown too that $q_{1}=p_{3}\equiv 0$.

Note that, while $\beta $ is the actual propagation constant for the FF
component of the soliton-carrying beam, $\beta _{1}$ and $2\beta _{2}$ are
the wave numbers of linear plane waves propagating in the medium at the
frequencies $\omega $ and $2\omega $. In fact, $\beta $, which is different
from $\beta _{1}$ because of the nonlinearity and finite transverse size of
the soliton beam, is a parameter of the soliton family at fixed values of
the carrier-wave propagation constants $\beta _{1,2}$.

In addition to the differential equations (\ref{eqTE}), it also
follows from the
Maxwell's equations that the SH electric filed must satisfy a set of linear
algebraic relations, 
$d_{12m}a_{m}=d_{32m}a_{m}=d_{12m}b_{m}=d_{22m}b_{m}=d_{32m}b_{m}=0$. We do
not display details of analysis which shows that Eqs. (\ref{eqTE}) are
compatible with the extra linear relations at some special values of the
angles ($\varphi ,\vartheta $,$\gamma $). After lengthy transformations, in
all such cases Eqs. (\ref{eqTE}) can be reduced to a simple system of two
equations, 
\begin{eqnarray}
\frac{d^{2}R}{d\eta ^{2}} &=&R-uR+\delta \cdot R^{3},  \label{FF} \\
\frac{d^{2}u}{d\eta ^{2}} &=&4\frac{\beta ^{2}-\beta _{2}^{2}}{\beta
^{2}-\beta _{1}^{2}}\,u\pm R^{2},  \label{SH}
\end{eqnarray}
where $R$ and $u$ are certain combinations of the FF and SH electric fields, 
$\eta \equiv \sqrt{\beta ^{2}-\beta _{1}^{2}}x$, and $\delta $ is some other
constant.

In particular, a nontrivial compatibility case can be obtained by choosing $%
\cos \vartheta =\cos \gamma =\cos (2\varphi )=0$. One then has $d_{122}=\pm
d,$ $a_{2}=a_{3}=b_{1}=0$, and 
\begin{equation}
R=\sqrt{2}D\left( \beta /\beta _{2}\right) A,\,\,u=D(a_{1}+A^{2}/\varepsilon
_{2}),\,\,b_{3}=-\left( \sqrt{\beta ^{2}-\beta _{1}^{2}}/2\beta D\right) 
\frac{du}{d\eta },  \label{Ru}
\end{equation}
with $\delta \equiv \left( \beta ^{2}-\beta _{1}^{2}\right) /4\beta ^{2}$,$%
\;D\equiv 2d_{122}\omega ^{2}/c$.

A special case when all the equations are compatible and $\delta =0$ (no
effective cubic nonlinearity) corresponds (for instance) to $\cos
(2\vartheta )=\sin \varphi =\sin \gamma =0$, which again yields $d_{122}=\pm
d$. In this case, Eqs. (\ref{FF}) and (\ref{SH}) are equivalent to the
simplest version of the stationary $\chi ^{(2)}$ model, hence they have
well-known soliton solutions \cite{Jena}. If $\delta \neq 0$, the system
differs from the simplest one by the extra $\chi ^{(3)}$ term in Eq. (\ref
{FF}).

Soliton-generating models combining $\chi ^{(2)}$ and $\chi ^{(3)}$
nonlinearities are well known too, see, e.g., Refs. \cite{combined}.
Equations (\ref{FF}) and (\ref{SH}) are somewhat different from them, as
they contain the cubic term solely in the FF equation. In models of this
type, soliton solutions can be easily found \cite{combined}. A typical
example of a TE fundamental (single-humped) soliton obtained from Eqs. (\ref
{E}), (\ref{H}) and (\ref{Ru}) is displayed in Fig. 1.

Cases when all the equations are compatible and, eventually, amount to the
simplest system of Eqs. (\ref{FF}) and (\ref{SH}) are possible not only for
discrete sets of values of the three angles, as in the two examples shown
above, but also for certain one-parameter families, when two equations are
imposed on the three angles. Examples are
\begin{equation}
\sin ^{2}\varphi =2\sin ^{2}\gamma ,\;\sin ^{2}\vartheta \cos ^{2}\gamma =1;
\label{Family1}
\end{equation}
\begin{equation}
\cos ^{2}\varphi =2\sin ^{2}\gamma ,\;\cos \vartheta =\tan \varphi \cdot
\tan \gamma .  \label{Family2}
\end{equation}
These examples do not exhaust all the compatibility cases. A complete
analysis of the compatibility conditions for the TE type of the fields in
the cubic crystal proves to be very tedious.

\section{The TM case}

We now proceed to a more interesting TM model, when nonzero FF amplitudes
are $A_{1}\equiv A$,$\;A_{3}$, and $P_{2}\equiv P$, see Eqs. (\ref{E}) and (%
\ref{H}). For this model, stationary equations for the magnetic- and
electric-field amplitudes can be derived in a general form, which, however,
turns out to be very cumbersome. Therefore, we here concentrate on a
particular (but quite nontrivial) case, 
\begin{equation}
\sin 2\varphi = \sin \gamma = \cos 2\vartheta =0.  \label{def}
\end{equation}

In the present case, equations for nonvanishing fields can be cast into the
following form : 
\begin{eqnarray}
\beta P &=&-\beta _{0}\left( \varepsilon _{1}A+2dA_{3}b_{3}\right) , 
\nonumber \\
P^{\prime } &=&\beta _{0}[\varepsilon _{1}A_{3}+2d(Ab_{3}-A_{3}a_{1})], 
\nonumber \\
\beta _{0}P &=&-\beta A-A_{3}^{\prime },  \nonumber \\
\beta p_{2} &=&-\beta _{0}\left( \varepsilon _{2}a_{1}-dA_{3}^{2}\right) , 
\nonumber \\
p_{2}^{\prime } &=&2\beta _{0}\left( \varepsilon _{2}b_{3}+2dAA_{3}\right) ,
\nonumber \\
2\beta _{0}p_{2} &=&-2\beta a_{1}-b_{3}^{\prime }.  \label{equations}
\end{eqnarray}
It is convenient to define renormalized transverse ($E_{z}$,$e_{z}$) and
longitudinal ($E$,$e$) FF and SH electric fields, 
\begin{eqnarray}
\frac{2d\beta _{0}^{2}}{\sqrt{\beta ^{2}-\beta _{1}^{2}}\beta }\,A_{3}
&\equiv &-E_{z},\;\;\;\frac{2d\beta _{0}^{2}}{\beta ^{2}}A\equiv E,
\label{renormE} \\
\frac{2d\beta _{0}^{2}}{\sqrt{\beta ^{2}-\beta _{1}^{2}}\beta }\,b_{3}
&\equiv &-e_{z},\;\;\;\frac{2d\beta _{0}^{2}}{\beta ^{2}}a_{1}\equiv e.
\label{renorm_e}
\end{eqnarray}
Finally, the magnetic fields $P$ and $p_{2}$ can be eliminated from 
Eqs. (\ref{equations}), which yields a final fourth-order system for the
electric fields (where, again, $\eta \equiv 
\sqrt{\beta ^{2}-\beta_{1}^{2}}x$),
\begin{eqnarray}
\frac{dE}{d\eta } &=&E_{z}+Ee_{z}-(3\frac{\beta ^{2}}{\beta _{1}^{2}}-2\frac{%
\beta _{2}^{2}}{\beta _{1}^{2}})\,E_{z}e-\frac{\beta ^{2}-\beta _{1}^{2}}{%
\beta _{1}^{2}}\,E_{z}(E_{z}^{2}-e_{z}^{2}),  \nonumber \\
\frac{dE_{z}}{d\eta } &=&E-E_{z}e_{z},  \label{EE} \\
\frac{de}{d\eta } &=&2e_{z}+\frac{3\beta ^{2}-\beta _{1}^{2}}{\beta _{2}^{2}}%
\,EE_{z}-\frac{\beta ^{2}-\beta _{1}^{2}}{\beta _{2}^{2}}\,E_{z}^{2}e_{z}, 
\nonumber \\
\frac{de_{z}}{d\eta } &=&2\frac{\beta ^{2}-\beta _{2}^{2}}{\beta ^{2}-\beta
_{1}^{2}}\,e+E_{z}^{2}\,.  \label{ee}
\end{eqnarray}
This model with mixed $\chi ^{(2)}$ and $\chi ^{(3)}$ nonlinearities appears
to be novel, as compared to various known models of the 
$\chi ^{(2)}$-$\chi^{(3)}$ type \cite{combined}.

It is relevant to stress that, since both the TE and TM models, composed of
Eqs. (\ref{FF}), (\ref{SH}) and (\ref{EE}), (\ref{ee}), respectively, have
been derived directly from the Maxwell's equations, without resorting to the
paraxial approximation \cite{Jena}, the applicability of these models is not
limited to the case of broad solitons, whose width is much larger than the
carrier wavelength. In other words, the final equations derived above apply
as well to very narrow {\it subwavelength} solitons, which have recently
attracted considerable attention in spatial and temporal models with the
Kerr nonlinearity \cite{sub}, but were not thus far considered in 
$\chi^{(2)}$ systems.

\section{TM\ solitons}

As the system of equations (\ref{EE}) and (\ref{ee}) is essentially
different from the standard models, it is necessary to investigate its
soliton solutions, which we have done by means of the usual shooting
technique. Fundamental and higher-order solitons were found in a broad
parametric region.

The first noteworthy result is that fundamental solitons are characterized
by a single-humped profile of the {\em longitudinal} electric FF field,
corresponding to an even single-humped function $E_{z}(x)$, while the
transverse FF field is described by an {\em odd} function $E(z)$. In other
words, only the distribution of the squared longitudinal electric FF field, $%
E_{z}^{2}(x)$, is single-humped in the fundamental TM soliton, while the
distribution of the squared transverse FF field, $E^{2}(x)$, is {\em %
double-humped}, with $E^{2}(0)=0$. As for the SH components, it is evident
from Eqs. (\ref{ee}) that the corresponding transverse and longitudinal
fields $e$ and $e_{z}$ must be, respectively, even and odd functions of $x$,
whatever parity of the FF fields $E$ and $E_{z}$. This feature is confirmed
by Fig. 2; however, a nontrivial property of the fundamental soliton is that
the even function $e(x)$ is {\em not} monotonic in the interval $0<x<\infty $%
, giving rise to three humps and two zeros in the distribution of the
corresponding squared field $e^{2}(x)$.

Note that the opposite parities of the FF transverse and longitudinal fields 
$E(x)$ and $E_{z}(x)$ is an obvious consequence of Eqs. (\ref{EE}).
Evidently, besides the structure seen in Fig. 2, with the odd transverse and
even longitudinal FF fields, the latter condition is also compatible with
the opposite case, when $E(x)$ and $E_{z}(x)$ would be, respectively, even
and odd. However, numerical solutions have never turned up TM solitons of
that type.

Lastly, Fig. 2 clearly suggests a conclusion which is also supported by
numerical solutions obtained at many other values of the parameters: in
terms of the renormalized variables defined by Eqs. (\ref{renormE}) and (\ref
{renorm_e}), the maximum values of the longitudinal and transverse fields
are nearly equal. To compare the maximum values of the physical fields, one
should undo the rescalings (\ref{renormE}) and (\ref{renorm_e}). An obvious
result is that the amplitudes of the longitudinal FF and SH electric fields
remain much smaller than the amplitudes of the transverse fields if $\sqrt{%
\beta ^{2}-\beta _{1}^{2}}\ll \beta $, which is the case for the usual
solitons whose width is much larger than the carrier wavelength. For narrow
solitons, whose width is comparable to or smaller than the wavelength, the
physical amplitudes of the longitudinal and transverse fields become
comparable too.

Thus, the properties of the TM soliton are drastically different from, and
essentially more nontrivial than those of the TE soliton. Indeed, the latter
one has the transverse FF electric field described by an even function $%
A_{2}(x)$, which is monotonically decaying in the interval $0<x<\infty $.
Another strong difference is that, although the SH transverse electric field
in both the TM and TE solitons is described by an even function, only in the
latter case this function is monotonic at $0<x<\infty $.

It is also natural to compare the TE and TM solitons found above with
solitons in the usual $\chi ^{(2)}$ models, which, as a matter of fact, are
also of the TE type. As well as the TE soliton shown in Fig. 1, fundamental
solitons in the usual models are characterized by even distributions of both
FF and SH electric fields, with the single hump at the center of the
soliton. Thus, the fundamental TM solitons found in this work are
drastically different from the traditional $\chi ^{(2)}$ solitons.

The numerical integration of Eqs. (\ref{EE}) and (\ref{ee}) readily
generates, alongside the fundamental TM solitons, higher-order ones. A
typical example of a second-order TM soliton, in which each component has an
additional extremum inside the interval $0<x<\infty $, is displayed in Fig.
3. It is quite easy to find solitons of still higher orders too.

It is relevant to mention that higher-order solitons are well known in usual 
$\chi ^{(2)}$ models, both in the simplest two-component systems (see, e.g.,
Refs. \cite{higher})\ and in more sophisticated three- \cite{Alan} and
four-component \cite{Bragg}\ models. In most cases, higher-order solitons
are subject to strong dynamical instability. Nevertheless, they were found
to be (numerically) stable in three- and four-component models combining the 
$\chi ^{(2)}$ nonlinearity and resonant reflections on a Bragg grating \cite
{Bragg}. Direct analysis of the soliton stability in the present system is
quite difficult and is beyond the scope of this work, as it is necessary to
simulate the full system of the Maxwell's equations. Nevertheless, it seems
very plausible that both the TE and TM fundamental solitons may be stable,
while the higher-order ones are unstable.

\section{Conclusion}

We have derived, starting from the Maxwell's equations and without resort to
the paraxial approximation, equations describing stationary
(1+1)-dimensional beams propagating at an arbitrary direction in an optical
crystal with cubic symmetry and purely quadratic $\chi ^{(2)}$ nonlinearity.
The equations were derived separately for beams of the TE and TM types. In
both cases, they contain $\chi ^{(2)}$ and $\chi ^{(3)}$ nonlinear terms,
the latter ones being generated via the cascading mechanism. The final TE
equations and soliton solutions to them are quite similar to those in
previously known models with mixed $\chi ^{(2)}$-$\chi ^{(3)}$
nonlinearities. On the contrary to this, the TM model turns out to be very
different from the previously known ones. It consists of four first-order
equations for transverse and longitudinal components of the electric field
at the fundamental and second frequencies. Fundamental-soliton solutions to
the TM model are also drastically different from the usual $\chi ^{(2)}$
solitons, in terms of the parity of their components. Most noteworthy, the
transverse and longitudinal components of the electric field at the
fundamental frequency in these solitons have, respectively, odd and
single-humped even shapes. Besides that, it was also concluded that maximum
values of the longitudinal and transverse fields become comparable for very
narrow solitons, whose width is commensurate to or smaller than the carrier
wavelength.

\newpage

\section*{Figure Captions}

Fig. 1. A typical example of a bright spatial soliton with the TE
polarization at values of the parameters $\left( \beta ^{2} -
\beta_{2}^{2}\right) /\left( \beta ^{2}-\beta_{1}^{2}\right) = 0.8$ and
$\delta = 0.1$. Shown in this figure are the fields $R$ and $u$, obtained
from Eqs. (7) and (8), vs. the normalized transverse coordinate $\eta $.

Fig. 2. A typical example of a fundamental bright spatial soliton with the
TM polarization. Here and in the subsequent figure, the continuous and
dashed lines show, respectively, the transverse and longitudinal
normalized fields: (a) the FF fields $E$ and $E_{3}$, and (b) the SH
fields $e$ and $e_{3}$. 
The parameters are $\left( \beta _{1}/\beta \right) ^{2}=0.60$, 
$\left(\beta _{2}/\beta \right) ^{2}=0.80$.

Fig. 3. A typical example of a second-order bright spatial soliton with the
TM polarization. The parameters are $\left( \beta _{1}/\beta \right)
^{2}=\left( \beta _{2}/\beta \right) ^{2}=0.35$.

\end{document}